\documentstyle[sprocl]{article}
\def\be{\begin{equation}}
\def\ee{\end{equation}}
\def\bea{\begin{eqnarray}}
\def\eea{\end{eqnarray}}

\newcommand{\p}[1]{(\ref{#1})}

\begin{document}
\begin{flushright}
Preprint DFPD 99/TH/54 \\
hep-th/9912264
\end{flushright}

\bigskip
\title{THE $OSp(1|4)$ SUPERPARTICLE AND EXOTIC BPS STATES\footnote{Talk given
by D.S. at the XIV-th Max Born Symposium, Karpacz, Poland, 
September 21--25, 1999.}}
\author{Igor Bandos}
\address{Institute for Theoretical Physics, National Science Center 
``Kharkov Institute of Physics and Technology'',
 Kharkov, 61108, Ukraine}
\author
{Jerzy Lukierski\footnote{Supported in part by { KBN} grant
{ 2P03B13012}}} 
\address
{Institute for Theoretical Physics 
University of Wroclaw, \\
50-204 Wroclaw, Poland}
\author{and}  
\address{}
\author{Dmitri Sorokin
\footnote{On leave from Kharkov Institute of Physics and
Technology, Kharkov, 61108, Ukraine}}
\address{INFN, Sezione Di Padova\\
Via F. Marzolo, 8, 35131 Padova, Italia}
%
%

\maketitle
\abstracts{We discuss the dynamics of a superparticle in a superspace whose
isometry is generated by the superalgebra $OSp(1|4)$ or its central--charge 
contraction. Extra coordinates of the superspace associated with tensorial
central charges are shown to describe spin degrees of freedom of the superparticle,
so quantum states form an infinite tower of (half)--integer helicities. A peculiar
feature of the model is that it admits BPS states which preserve 3/4 of
target--space supersymmetries.}

In this contribution we present results of work done on studying superparticle
models whose symmetry and physical properties are defined by a supersymmetric
algebra extended by tensorial charges \cite{bl,bls,blps}.

A motivation for this study has been to understand the physical meaning of
tensorial ``central'' charges when they are associated not with superbranes but
with relativistic (point--like) particles.

As we shall see, in the case of superparticles tensorial central charges have
different physical meaning than that of superbranes. They correspond to spin
degrees of freedom of the superparticles, while it is well known that in the
case of branes tensorial charges describe the coupling of branes to tensor 
gauge fields of target--space supergravity. That is, brane tensorial
charges are similar to electric and magnetic charges of particles.

For instance, a membrane in $D=4$ or $D=11$ target space couples to a
three--form gauge field $A_{mnp}(x)$. In the membrane worldvolume action 
\cite{bst} this coupling is described by the Wess--Zumino term
\begin{equation}\label{1}
S_{coupl}=\int_{{\cal M}_3}dx^mdx^ndx^pA_{mnp}(x),
\end{equation} 
where the integral is taken over the membrane worldvolume ${\cal M}_3$ of 
the pull back of the gauge field. As was shown in \cite{agit}, the membrane
charge associated with this coupling is a two--rank tensor of the form
\begin{equation}\label{2}
Z^{mn}=\int_{{\cal M}_2}dx^m\wedge dx^n,
\end{equation}   
where the integral is taken over the two--dimensional surface of the membrane.

Using the supermembrane action \cite{bst} the authors of \cite{agit} derived
the form of the superalgebra generated by the Noether charges of the 
supermembrane. The supertranslation algebra thus obtained was shown to contain
the contribution of the membrane charge (\ref{2}) to the r.h.s. of the 
anticommutator of the fermionic supercharges
\begin{equation}\label{3}
\{Q_\alpha,Q_\beta\}=-2(C\gamma_m)_{\alpha\beta}P^m
+(C\gamma_{mn})_{\alpha\beta}Z^{mn},
\end{equation}
where $P_m$ is the standard momentum (or bosonic translation) generator.

Thus, the worldvolume actions for the superbranes imply that the underlying
supersymmetry of superbranes is described by a superalgebra extended by 
tensorial central charges \cite{agit,st,hammer}.

Recent analysis carried out in \cite{craps,fp} has demonstrated that when 
superbranes propagate in anti--de--Sitter superbackgrounds, the superalgebra
of their Noether supercharges gets extended to a corresponding maximal
$OSp$ superalgebra.

For instance, for branes in $D=10$ and $D=11$ target superspaces with $AdS$
geometry the Noether charges generate the $OSp(1|32)$ superalgebra.
In \cite{bars} it has been assumed that the underlying superalgebra of M--theory
should be even larger, namely, $OSp(1|64)$ which is the minimal simple 
superalgebra which contains the supertranslation algebra of M--theory 
with a two--form and a five--form central charge as a {\it subsuperalgebra}.

If the $OSp(1|32)$ and $OSp(1|64)$ supergroup are related to M--theory,
it seems instructive
to find and study simple dynamical systems whose properties would be governed
by $OSp(1|2n)$ supergroups. 
And this has been another motivation for our work.

To simplify consideration let us take the $OSp(1|4)$ supergroup which is the
isometry of $N=1,D=4$ superspace having the four--dimensional
$AdS_4$ space as a bosonic subspace (see \cite{blps} for the generic $OSp(1|2n)$
case). The $OSp(1|4)$ superalgebra has the following form
\begin{eqnarray}\label{2.1}
 [M_{ab},M_{cd}]&=& 
i(\eta_{ad}M_{bc} + \eta_{bc}M_{ad} 
- \eta_{ac}M_{bd} -\eta_{bd}M_{ac}),
\label{2.1aa}
\\ \cr
 [M_{ab},P_{c}]&= & i(\eta_{bc}P_{a}-\eta_{ac}P_{b}),
 \label{2.1b}
\\ \cr 
 [P_{a},P_{b}] &= & {i\over R^{2}} M_{ab},
 \label{2.1c}
\\ \cr 
 \left\{{Q}_{\alpha}, Q_{\beta}\right\}
&= & - 2(C\gamma^{a})_{\alpha\beta}P_{a}
  + {1\over R} \left(C\gamma^{ab}\right)_{\alpha\beta}
M_{ab},
\label{2.1d}
\cr
\cr  
 [M_{ab},Q_{\alpha}] &= &- {i\over 2} Q_{\beta}
\left(\gamma_{ab}\right)^{\beta}_{\ \alpha}, \quad
\gamma_{ab}={1\over 2}(\gamma_a\gamma_b-\gamma_b\gamma_a),
\label{2.1e}
\\ \cr
 [P_{a},Q_{\alpha}] & =&-
 {i\over 2R} Q_{\beta}
\left(\gamma_{a}\right)^{\beta}_{\ \alpha}.
\label{2.1f}
\end{eqnarray}
The generators $M_{ab}$ form the $SO(1,3)$ subalgebra \p{2.1aa},
and $M_{ab}$ and $P_a$ (a=0,1,2,3) generate the $SO(2,3)\sim Sp(4)$ 
subalgebra of $OSp(1|4)$.
$Q_\alpha$ are four Majorana spinor generators of $OSp(1|4)$.
The parameter $R$ is the $AdS_4$ radius, and $C_{\alpha\beta}$ is
a charge conjugation $Sp(4)$--invariant matrix. 

The $N=1, D=4$ supertranslation algebra with tensorial central charges
(\ref{3}) is a contraction of the $OSp(1|4)$ superalgebra carried out in the 
following way. One rescales the $SO(1,3)$ generators as 
\begin{equation}\label{211}
M_{ab}=RZ_{ab} 
\end{equation}
and
takes the limit $R\rightarrow\infty$, which corresponds to the limit where
$AdS_4$ becomes assimptotically flat. Then $Z_{ab}$ and $P_{a}$ become Abelian
and commute with $Q_{\alpha}$, and the anticommutator of $Q_{\alpha}$ reduces
to that in (\ref{3}).

Consider now a simple ``oscillator'' realization of $OSp(1|4)$. For this let
us combine $P_{a}$ and $M_{ab}$ into a symmetric spin--tensor generator
\begin{equation}\label{4}
M_{\alpha\beta}=-2(C\gamma_a)_{\alpha\beta}P^a
+(C\gamma_{ab})_{\alpha\beta}Z^{ab}
\end{equation}
so that
\begin{eqnarray}\label{5}
 && [M_{\alpha\beta}, M_{\gamma\delta}]=-{{4i}\over R}[C_{\gamma (\alpha } 
M_{\beta ) \delta}
 + C_{\delta (\alpha } M_{\beta )\gamma}],\qquad \nonumber \\ 
 && [ M_{\alpha\beta }, Q_\gamma  ] =-{{4i}\over R} C_{\gamma ( \alpha  } 
Q_{\beta )}, \quad 
\{ Q_{\alpha}, Q_\beta \} = M_{\alpha\beta}. 
\end{eqnarray}
If we introduce a Grassmann--even spinor operator $\lambda_\alpha$, which
forms the Heisenberg algebra
\begin{equation}\label{6}
[\lambda_\alpha,\lambda_\beta]={{2i}\over{R}}C_{\alpha\beta},
\end{equation}
and a Grassmann--odd scalar $\psi$ such that
\begin{equation}\label{7}
\psi^2={1\over 2}, \quad \psi\lambda_\alpha-\lambda_\alpha\psi=0,
\end{equation}
we can realize $M_{\alpha\beta}$ and $Q_\alpha$ as follows
\begin{equation}\label{8}
M_{\alpha\beta}={1\over 2}(\lambda_\alpha\lambda_\beta
+\lambda_\beta\lambda_\alpha), \quad Q_\alpha= \lambda_\alpha\psi.
\end{equation}
Using the commutation relations for $\lambda_\alpha$ and $\psi$ it is not
hard to check that $Q_\alpha$ and $M_{\alpha\beta}$ represented in this way
generate the $OSp(1|4)$ superalgebra, ${\cal Z_A}\equiv (\lambda_\alpha,\psi)$
forming the fundamental representation of  $OSp(1|4)$.

Note that in the limit $R\rightarrow\infty$, $\lambda_\alpha$ become commuting
quantities (see (\ref{6})). Then we observe that
\begin{equation}\label{9} 
P_a=\lambda\gamma_a\lambda  
\end{equation}
in which one recognizes the Cartan--Penrose relation implying that
\begin{equation}\label{10}
P_aP^a\equiv 0,
\end{equation}
due to $D=4$ $\gamma$--matrix identities. We thus assume that at 
$R\rightarrow\infty$ the ``oscillator'' realization of $OSp(1|4)$ may 
correspond to a massless $D=4$ superparticle with $\lambda_\alpha$ playing
the role of a twistor--like variable.

Before contraction such a superparticle propagates in the supergroup manifold
$OSp(1|4)$ parametrized by coordinates $x^a,y^{ab}$ and $\theta^\alpha$ 
associated, respectively, with the generators $P_a,M_{ab}$ and $Q_\alpha$.

An action for this superparticle can be constructed in a way similar to that used
for the first time by Ferber \cite{f} for developing the supertwistor 
formulation of supersymmetric field theories.

In our case the superparticle worldline 
$Z^M(\tau)=(x^a(\tau),y^{ab}(\tau),$
$\theta^\alpha(\tau))$ on the $OSp(1|4)$ manifold is parametrized by the time 
variable $\tau$. To construct the action we pick the worldline pull back of 
left--invariant $OSp(1|4)$ Cartan one--forms taking values in the $SO(2,3)$ 
subalgebra of $OSp(1|4)$
\begin{equation}\label{11}
\Omega^{\alpha\beta}(Z)=dZ^M(\tau)\Omega_M^{\alpha\beta}(Z)
\end{equation}
and contract the spinor indices with {\it commuting} spinor variables 
$\lambda_\alpha(\tau)$ which are classical counterparts of the spinor operators
(\ref{6}) used to realize the generators (\ref{8}) of  $OSp(1|4)$. $\lambda_\alpha$ 
become non--commutative upon solving for second--class constraints of the model
and passing from Poisson to Dirac brackets.

The action we thus obtain has the twistor--like form
\begin{equation}\label{12}
S=\int d\tau\lambda_\alpha\lambda_\beta \partial_\tau Z^M\Omega_M^{\alpha\beta}.
\end{equation}  
By construction (\ref{12}) is invariant under the $OSp(1|4)$ transformations
of the coordinates $Z^M=(x^a,y^{ab},\theta^\alpha)$ and under the $SO(2,3)$ subgroup
of $OSp(1|4)$ acting in the tangent space of the supermanifold $OSp(1|4)$.
To better understand the symmetry structure of the action (\ref{12}) we note
that the isometry of the supergroup manifold $OSp(1|4)$ is the direct product
of two supergroups $OSp(1|4)_L\times OSp(1|4)_R$. By using in (\ref{12})
 the $OSp(1|4)$--left--invariant Cartan forms corresponding only to $SO(2,3)_R$
we break ``right--acting'' $OSp(1|4)_R$ down to $SO(2,3)$.

To be able to analyze the action (\ref{12}) one should know an explicit expression
for the Cartan forms $\Omega^{\alpha\beta}$. This can be obtained by substituting
an appropriate parametrization of the  $OSp(1|4)$ supergroup element $G(Z)$ into
the definition of the Cartan forms
\begin{equation}\label{13}
\Omega=-iG^{-1}dG\equiv \Omega^{\alpha\beta}(Z)M_{\alpha\beta}+E^\alpha(Z) Q_\alpha,
\end{equation}
where $E^\alpha$ is a spinorial Cartan form associated with the supercharge generators.

In \cite{blps} we have found a parametrization of $G(Z)$ 
which allowed us to obtain simpler 
expressions for the Cartan forms of $OSp(1|4)$, and generically of $OSp(1|2n)$,
than those derived in earlier papers \cite{z,gm,is}. We have got
\begin{equation}\label{14}
\Omega^{\alpha\beta}=v^\alpha_{~\gamma}(y)v^\beta_{~\delta}(y)
\left [(\gamma_a)^{\gamma\delta}e^a(x)+R(\gamma_{ab})^{\gamma\delta}\omega^{ab}(x) + 
\theta^{(\gamma}{\cal D}\theta^{\delta)}+(dvv^{-1})^{\gamma\delta}\right],
\end{equation}
where $v^\alpha_{~\gamma}(y)$ are $SO(1,3)$ matrices in the spinor representation,
$e^a(x)$ and $\omega^{ab}(x)$ are, respectively, the vierbein and the spin 
connection on the bosonic coset space $AdS_4={{SO(2,3)}\over{SO(1,3)}}$, and 
${\cal D}=d+{1\over 2}\omega^{ab}(x)\gamma_{ab}+{1\over{2R}}e^a(x)\gamma_a$ is the
covariant $AdS_4$ differential.

Substituting (\ref{14}) into the superparticle action (\ref{12}) and making the 
redefinition of $\lambda_\alpha(\tau)$
\begin{equation}\label{15}
\Lambda_\alpha(\tau)=\lambda_\beta v^\beta_{~\alpha}(y),
\end{equation}
we rewrite the action in the following form
\begin{equation}\label{16}
S=\int_{{\cal M}_\tau}\left[\Lambda\gamma_a\Lambda(e^a-i\theta\gamma^a{\cal D}\theta)
+\Lambda\gamma_{ab}\Lambda\left(R\omega^{ab}
+{i\over 2}\theta\gamma^{ab}{\cal D}\theta+tr(dvv^{-1}\gamma^{ab})\right)\right].
\end{equation}
Note that the $SO(1,3)$ coordinates $y^{ab}$ enter this action only through the
last term $dvv^{-1}$. All other terms depend on the bosonic $x^a$ and fermionic
$\theta^\alpha$ coordinates of the coset superspace ${{OSp(1|4)}\over{SO(1,3)}}$
whose bosonic subspace is $AdS_4$.

If in (\ref{16}) we skip the terms with $\Lambda\gamma_{ab}\Lambda$ we will get
the action describing a massless superparticle propagating in the $AdS_4$ superspace
\cite{blps}.

On the other hand, if we take the limit $R\rightarrow\infty$ then, 
as we have discussed above, the $OSp(1|4)$ superalgebra gets contracted to the 
super--Poincare algebra with the tensorial central charge, and the $OSp(1|4)$ 
superparticle action reduces to the action  which describes a superparticle 
propagating in $N=1, D=4$ flat superspace $(x^a,\theta^\alpha)$ extended by
tensorial coordinates $y^{ab}$  associated now with central charge momentum 
generators $Z_{ab}$ (\ref{211}). The superparticle action takes the form
\begin{equation}\label{17}
S=\int \left[\Lambda\gamma_a\Lambda(dx^a-i\theta\gamma^ad\theta)
+\Lambda\gamma_{ab}\Lambda(dy^{ab}+i\theta\gamma^{ab}d\theta)\right].
\end{equation}

This action, which by construction obeys supersymmetry with tensorial charges, was
proposed in \cite{bl}. The massless $N=1, D=4$ superparticle described by 
this action possesses quite unusual features.

One of them is that the action is invariant under fermionic (so called 
$\kappa$--symmetry) transformations with {\it three} independent parameters. The 
$\kappa$--symmetry transformations have the following form
$$
\delta_\kappa\theta^\alpha=\kappa^I(\tau)\mu^\alpha_I \quad (I=1,2,3)
$$
\begin{equation}\label{18}
\delta_\kappa x^a=i\theta\gamma^a\delta_\kappa\theta, \quad
\delta_\kappa y^{ab}=-i\theta\gamma^{ab}\delta_\kappa\theta,
\end{equation}
where $\mu^\alpha_I$ are three linearly independent commuting spinors orthogonal
to $\lambda_\alpha$, i.e. $\lambda_\alpha\mu^\alpha_I=0.$

Remember, that standard superparticle and, in general, superbrane actions
are invariant under $\kappa$--symmetry transformations with the number of 
independent parameters  being half the number of the spinor coordinates
of target superspace. 
So in $N=1, D=4$ superspace the standard massless 
superparticle is invariant under {\it two} independent $\kappa$-symmetries. 
As was realized in \cite{stv}, in the twistor--like formulation 
$\kappa$-symmetries can be made irreducible and traded for $n=2$ 
extended worldline supersymmetry with transformation properties of
$\theta^\alpha$ and $x^a$ being
\begin{equation}\label{19}
\delta\theta^\alpha=\epsilon_1(\tau)\lambda^\alpha
+\epsilon_2(\tau)(\gamma_5\lambda)^\alpha, \quad \delta x^a=\theta\gamma^a\delta\theta.
\end{equation}

One can easily observe the difference  between the transformations (\ref{18})
and (\ref{19}). The latter explicitly contain $\lambda^\alpha$, while the
former involve spinors orthogonal to $\lambda^\alpha$.

The invariance of the superbrane action with respect to local fermionic 
transformations implies that there exist superbrane configurations which preserve
the number of target space supersymmetries which is equal to or less than 
the number of independent $\kappa$--symmetries. Such supersymmetric 
states saturate the Bogomol'nyi--Prassad--Sommerfeld energy bound. 
Thus in the case of the standard superbranes the number of unbroken target--space
supersymmetries is not higher than ${1\over 2}$ supersymmetry of the target--space
vacuum.

For instance, in the case of the standard $N=1, D=4$ massless superparticle only two
of four target--space supersymmetries are unbroken. While in the case of the 
superparticle with tensorial central charges there are BPS superparticle 
configurations with {\it three} target--space supersymmetries, i.e. $3\over 4$ 
of supersymmetry remain unbroken \cite{bl}. 

Recently the possibility of the existence of exotic BPS brane configurations
preserving more than $1\over 2$ supersymmetry has been discussed in \cite{gh}. 
The superparticle model based on the action (\ref{17}) is an example of such
configurations.

This unusual property is also characteristic of the superparticle propagating on 
the whole supergroup manifold $OSp(1|4)$ described by the action (\ref{16}) 
\cite{blps}.  The algebraic reason for such an exotic situation is that, as we
have discussed above, the $OSp(1|4)$ superalgebra and its central charge 
contraction is realized in such a way that the anticommuting Poisson brackets 
of the Noether supercharges $Q_\alpha$ derived from (\ref{16}) 
is equal to the $\lambda_\alpha$--spinor belinears
\begin{equation}\label{20}
\{Q_\alpha,Q_\beta\}=M_{\alpha\beta}=\lambda_\alpha\lambda_\beta.
\end{equation}
The matrix $\lambda_\alpha\lambda_\beta$ is degenerate and has the rank one. 
Hence only one of four supergenerators $Q_\alpha$ (\ref{8}) has nonzero
anticommutator. To single out this supercharge, let us introduce a basis ($\mu^\alpha,
\mu^\alpha_I)$ $(I=1,2,3)$ in the spinor space such that 
({\rm compare with} (\ref{18}))
$$
\mu^\alpha \lambda_\alpha=1, \quad \mu^\alpha_I \lambda_\alpha=0.
$$
Than, in view of (\ref{7}) and (\ref{8}),
$$
Q=\mu^\alpha Q_\alpha=\psi \quad \Rightarrow \quad Q^2={1\over 2}
$$
corresponds to one broken supersymmetry and three supercharges
$$
Q_I=\mu^\alpha_IQ_\alpha\quad \Rightarrow \quad \{Q_I,Q_J\}=\{Q_I,Q\}=0
$$
anticommute with themselves and with $Q$. Hence, $Q_I$ act trivially on BPS superparticle
states and correspond to three unbroken supersymmetries.

Another feature of the superparticle model with the tensorial charge coordinates is
the physical meaning of these extra variables. As the Hamiltonian analysis and the 
quantization of this superparticle model have shown \cite{bls}, only one of the
six tensorial charge coordinates $y^{ab}$ is independent due to a large number
of constraints. This coordinate takes discrete integer values $n=2s$ and labels
half--integer and integer helicities $(s=0,\pm {1\over 2}, \pm 1,..., \infty)$ of 
massless quantum states of the superparticle in $D=4$ space--time. 

Let us consider this in more detail. In the Weyl representation of spinors
$$
\lambda_\alpha=(\lambda_A,\bar\lambda^{\dot A}), \quad 
\theta_\alpha=(\theta_A,\bar\theta^{\dot A}), \quad A=1,2; \quad \dot A=1,2
$$
the action (\ref{17}) takes the form
\begin{equation}\label{20a}
S=\int\left[\lambda_A\bar\lambda_{\dot A}(dx^{A\dot A}-i\theta^Ad\bar\theta^{\dot A}
+id\theta^A\bar\theta^{\dot A})+\lambda_A\lambda_B(dy^{AB}-i\theta^Ad\theta^{A})+c.c.
\right],
\end{equation}
where
$$
x^{A\dot A}=x^a\sigma_a^{A\dot A}, \quad y^{AB}=y^{ab}(\sigma_{ab})^{AB},\quad
y^{\dot A\dot B}=y^{ab}(\bar\sigma_{ab})^{\dot A\dot B},
$$
and $\sigma_a^{A\dot A}$ are the Pauli matrices $(\sigma_{ab})_A{}^B =
1/2i \left(\sigma_{a A\dot{B}} \tilde{\sigma}_b^{\dot{B}B} -
( a \leftrightarrow b )\right)$.

 From (\ref{20a}) we obtain that the canonical momenta associated 
with $D=4$ coordinates
$x^a$ and tensorial charge coordinates $y^{ab}$ are, respectively,
\begin{equation}\label{20b}
p_{A\dot A}=\lambda_A\bar\lambda_{\dot A}
\end{equation}
$$
Z_{AB}=\lambda_A\lambda_B,
\quad Z_{\dot A\dot B}=\bar\lambda_{\dot A}\bar\lambda_{\dot B}.
$$
Notice, in particular, that $p^{A\dot A}p_{A\dot A}\equiv 0$, and, 
hence, the superparticle is massless. 

The Cartan--Penrose relation (\ref{20b}) establishes the correspondence between
three independent components of lightlike $p_a=\sigma_a^{A\dot A}p_{A\dot A}$ and
three components of $\lambda_A,\bar\lambda_{\dot A}$. Only the phase of $\lambda$
\begin{equation}\label{20c}
\lambda_A=e^{i\varphi(\tau)}\lambda^0_A
\end{equation}
remains undetermined.

If $y^{ab}=0$, we deal with a twistor superparticle considered in \cite{f,shi},
its action being
\begin{equation}\label{20d}
S=\int\lambda_A\bar\lambda_{\dot A}(dx^{A\dot A}-i\theta^Ad\bar\theta^{\dot A}
+id\theta^A\bar\theta^{\dot A}).
\end{equation}
In addition to all symmetries of the  action (\ref{20a}), the action (\ref{20d}) is
invariant under local $U(1)$ transformations
\begin{equation}\label{u1}
\lambda_A~~\rightarrow~~e^{i\varphi(\tau)}\lambda_A, \quad
\bar\lambda_{\dot A}~~\rightarrow~~e^{-i\varphi(\tau)}\bar\lambda_{\dot A}.
\end{equation}
This gauges away the phase component of $\lambda_A$ and establishes the 
one--to--one correspondence between the independent 
components of the twistor superparticle momentum $p_a$ and $\lambda_A$.

As the quantization of the action (\ref{20d}) has shown \cite{f,shi}, the quantum
states of the $N=1, D=4$ twistor superparticle form chiral supermultiplets
of physical states with helicity $0$ and ${1\over 2}$. These supermultiplets are
described by chiral superfields $\Phi(x^a-i\theta\sigma^a\bar\theta,\theta_A)$.

In the case of the superparticle model (\ref{20a}) with additional tensorial 
coordinates $y^{ab}$ there is no local $U(1)$ symmetry (\ref{u1}). The compact phase
component of $\lambda_A$ becomes a {\it physical} momentum degree of freedom which
corresponds to a single independent component of 
tensorial charge momenta $Z_{ab}$.

So  the superparticle wave function in the momentum representation now becomes
\begin{equation}\label{21a}
\Phi(p_a,\varphi,\theta^\alpha)
=\Sigma^\infty_{n=0}\left[e^{in\varphi}\Phi_n(p_a,\theta^\alpha)+
e^{-in\varphi}\bar\Phi_n(p_a,\theta^\alpha)\right ].
\end{equation}
Integer $n$ is associated with an independent discrete (or quantized) 
component of the
central charge coordinates $y^{ab}$, which is the Fourier 
image of the compact phase
momentum component $\varphi$. This resembles a ``dual'' Kaluza--Klein effect when
instead of a spatial direction, 
compactified is the corresponding momentum coordinate
of the phase space.

In the Lorentz--covariant form the first--quantized
wave function of the superparticle in the momentum representation looks as follows
\cite{bls}
\begin{equation}\label{21}
\Phi= \Sigma^\infty_{n=0} 
\Phi^{\alpha_1...\alpha_n}(p_m)
\lambda_{\alpha_1}...\lambda_{\alpha_n}+\chi
\Sigma^\infty_{n=0} 
\Psi^{\alpha_1...\alpha_n}(p_m)
\lambda_{\alpha_1}...\lambda_{\alpha_n},
\end{equation}
where $\chi=\theta^\alpha\lambda_\alpha, \chi^2\equiv 0$.
Each component of this series forms an irreducible 
representation of the Lorentz group
$SO(1,3)$ and describes a massless 
state with integer or half--integer superhelicity 
depending whether $n$ is even or odd.

Thus, the superparticle model with tensorial charges produces, 
upon quantization, an
infinite tower of massless fields of higher spin. 
The structure of the wave function which
describes higher--spin states is similar to that used in the formalism developed by
Vasiliev (for a recent review see \cite{misha}) 
to construct the theory of higher--spin fields. Hence, the model
which we have briefly described can be assumed to be a classical counterpart of the
field theory of higher spins.

To conclude, we have demonstrated that tensorial charges appearing in extensions of
supertranslation algebras may have different meaning than that one got accustomed to
in superbrane models. They may describe spinning degrees of freedom of a dynamical
system. Superparticle models on supergroup manifolds $OSp(1|2n)$ and their contractions
to flat superspaces with tensorial coordinates may give rise to exotic BPS 
configurations which preserve more than $1\over 2$ supersymmetry.

We have seen that the $AdS_4$ space is an intrinsic part of the construction of
the superparticle action (\ref{16}) on $OSp(1|4)$, which upon the contraction
describes free higher--spin states. It is well known that to switch on
interactions of higher--spin fields one needs the space--time to be of $AdS$ geometry
\cite{misha}.
So an interesting problem to study is whether the superparticle model on
$OSp(1|4)$ may help to make a progress in constructing the theory of interacting fields
of higher spin.

\end{document}